\documentclass[journal,10pt,letterpaper,final,twocolumn]{IEEEtran}%
\usepackage{amsmath}
\usepackage{amsfonts}
\usepackage{cite}
\usepackage{amssymb}
\usepackage{mathrsfs}
\usepackage{graphicx}
\usepackage[usenames,dvipsnames]{color}
\usepackage{epstopdf}
\usepackage[all]{xy}
\usepackage[]{mcode}
\usepackage{supertabular}
\usepackage{subcaption}
\usepackage{array}

\usepackage{algorithmic}
\usepackage{algorithm}

\providecommand{\U}[1]{\protect\rule{.1in}{.1in}}
\pdfoutput=1
\providecommand{\U}[1]{\protect\rule{.1in}{.1in}}

\newcommand{\qed}{\nobreak \ifvmode \relax \else
      \ifdim\lastskip<1.5em \hskip-\lastskip
      \hskip1.5em plus0em minus0.5em \fi \nobreak
      \vrule height0.75em width0.5em depth0.25em\fi}

\begin{document}

\title{Big Communications: Connect the Unconnected}
\author{Shuping Dang, \textit{Member, IEEE}, Chuanting Zhang, \textit{Member, IEEE}, \\Basem Shihada, \textit{Senior Member, IEEE}, and Mohamed-Slim Alouini, \textit{Fellow, IEEE}
  \thanks{© 2021 IEEE. Personal use of this material is permitted. Permission from IEEE must be obtained for all other uses, in any current or future media, including reprinting/republishing this material for advertising or promotional purposes, creating new collective works, for resale or redistribution to servers or lists, or reuse of any copyrighted component of this work in other works.
  
  The authors are with Computer, Electrical and Mathematical Science and Engineering Division, King Abdullah University of Science and Technology (KAUST), 
Thuwal 23955-6900, Saudi Arabia (e-mail: \{shuping.dang, chuanting.zhang, basem.shihada, slim.alouini\}@kaust.edu.sa).}}

\maketitle
\date{}

\begin{abstract}
In this article, we present the analysis of the digital divide to illustrate the unfair access to the benefits brought by information and communications technology (ICT) over the globe and provide our solution termed big communications (BigCom) to close the digital divide and democratize the benefits of ICT. To facilitate the implementation of BigCom, we give a complete framework of BigCom considering both technological and non-technological factors as well as a set of suggestions for content providers, mobile network operators, and governments. By implementing BigCom, we aim to connect the last four billion unconnected people living in far-flung and underdeveloped areas and achieve the goal of global and ubiquitous connectivity for everyone in the 6G era.
\end{abstract}

\section*{Introduction}
\IEEEPARstart{A}{s} fifth generation (5G) communication networks are being deployed since 2019, researchers of communications science in academia and industry have started paying attention to sixth generation (6G) communications and studying the application scenarios as well as the corresponding performance goals \cite{dang2020should,8412482,8808168}. However, with the rapid development of 5G information and communications technology (ICT), it is even difficult to believe that there are almost four billion people on the globe who are still unconnected or under-connected \cite{philbeck2017connecting}. 

Because of this surprising fact, researchers of ICT have reached a consensus that the benefits of ICT should be democratized over the world, rather than being reserved for urban and developed regions only. In this regard, different from 5G that is primarily dedicated to satisfying the application scenarios in populous urban areas with the dense deployment of telecommunication infrastructure, 6G aims to serve the far-flung regions with much less population, physical infrastructure, and limited financial resources and close the  digital divide in the 2030s \cite{9042251}. 

Accordingly, IEEE P2061, a standardization project initiated by IEEE, has been dedicated to enabling affordable broadband communications for rural areas \cite{9139048}. It is highly expected that the digital inclusion and accessibility over these underdeveloped areas will become the new economic impetus and greatly raise the levels of education, health care, and other public services. As a result, democratizing the benefits of ICT is the cornerstone of Goal 1 (No Poverty), Goal 4 (Quality Education), Goal 8 (Decent Work and Economic Growth), and, of course, Goal 9 (Industry, Innovation and Infrastructure) of the United Nations' Sustainable Development Goals (SDGs).

In this article, we analyze the causes of the four billion connected/under-connected population and propose a novel solution termed \textit{big communications} (BigCom). BigCom is anticipated to be a framework of democratizing the benefits of ICT and realizing global connectivity in the 6G era.

\section*{Human-Centric 6G Versus Digital Divide}
As proposed in our article titled `What should 6G be?' by \textit{Nature Electronics} \cite{dang2020should}, a unique trait of 6G communications compared to the previous generations is the human-centric nature. This human-centric perspective requires to pay attention to the real and heterogeneous needs of end users, rather than increasing certain performance indicators in a mindless manner. 

In metropolitan areas with a dense population, providing an extremely high throughput and a vast number of simultaneous connections with much higher standards of security, secrecy, and privacy is the key mission of 6G communications. On the other hand, in remote areas with only a few residents, basic data and telecommunication services should also be supported by 6G communications as a fundamental social utility, regardless of the profitability. This becomes another focus of 6G communications launched in the 2030s. Apart from conventional technical performance measures, it is also suggested by some researchers that the economic and sociological tools, e.g., the Gini index and the Lorenz curve, would be used to evaluate the fairness of access to the benefits of ICT among different areas \cite{6517050}.

It has been reported by the background paper to the special session of the Broadband Commission and the World Economic Forum at Davos Annual Meeting in 2017 that there was 53\% of the world's population that was still offline, which was about four billions of people in absolute numbers. Most of these offline people resided in Africa and Asia-Pacific, and there was 85\% of the offline population in the least developed countries, whereas the number was 22\% in developed countries, forming an evident \textit{digital divide}  \cite{philbeck2017connecting}.

\begin{figure*}[!t]
\centering
\includegraphics[width=5.5in]{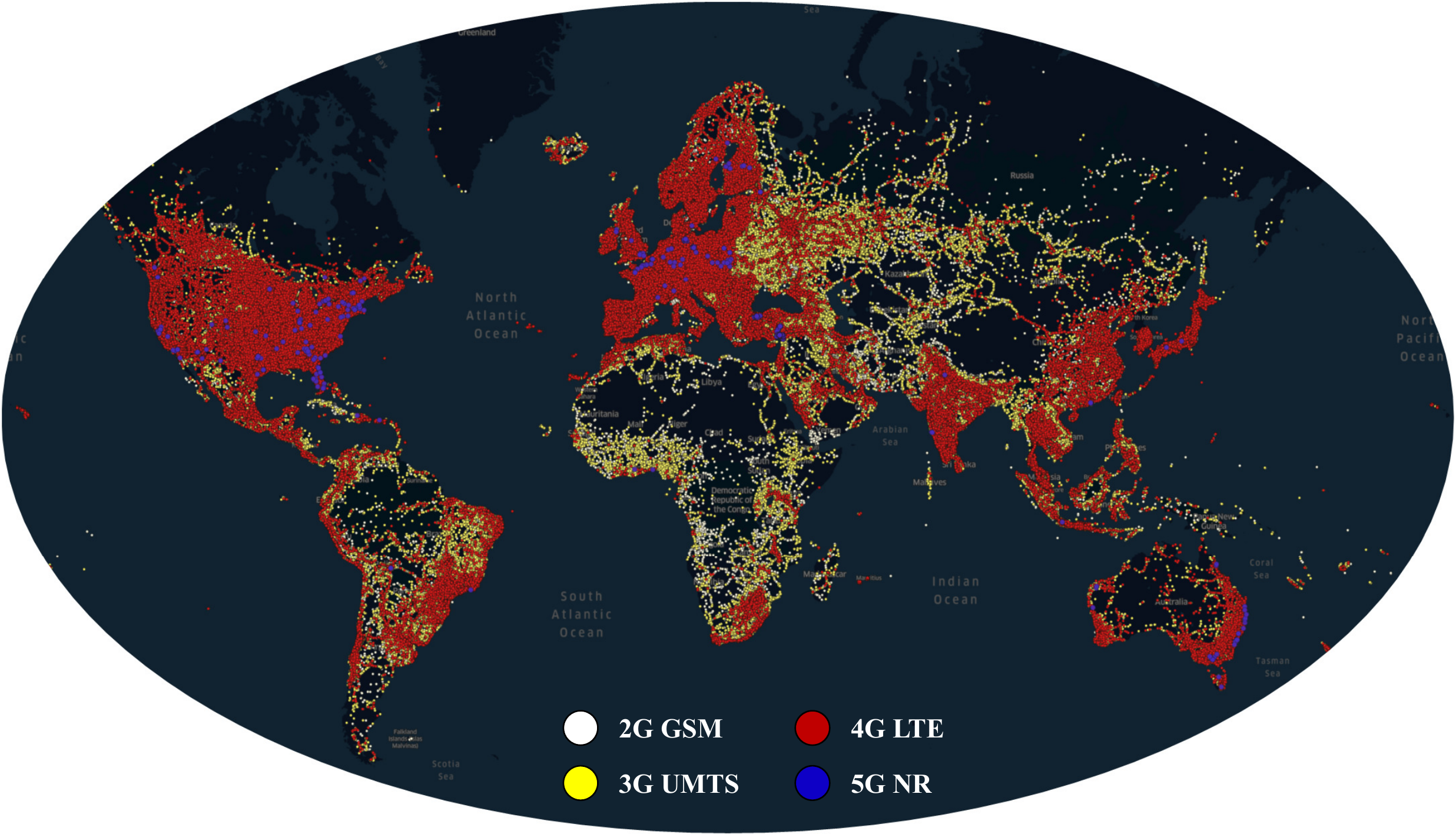}
\caption{Distribution of BSs all over the world: A visualization of the global digital divide.}
\label{bs_distribution}
\end{figure*}

\begin{figure}[!t]
\centering
\includegraphics[width=3.5in]{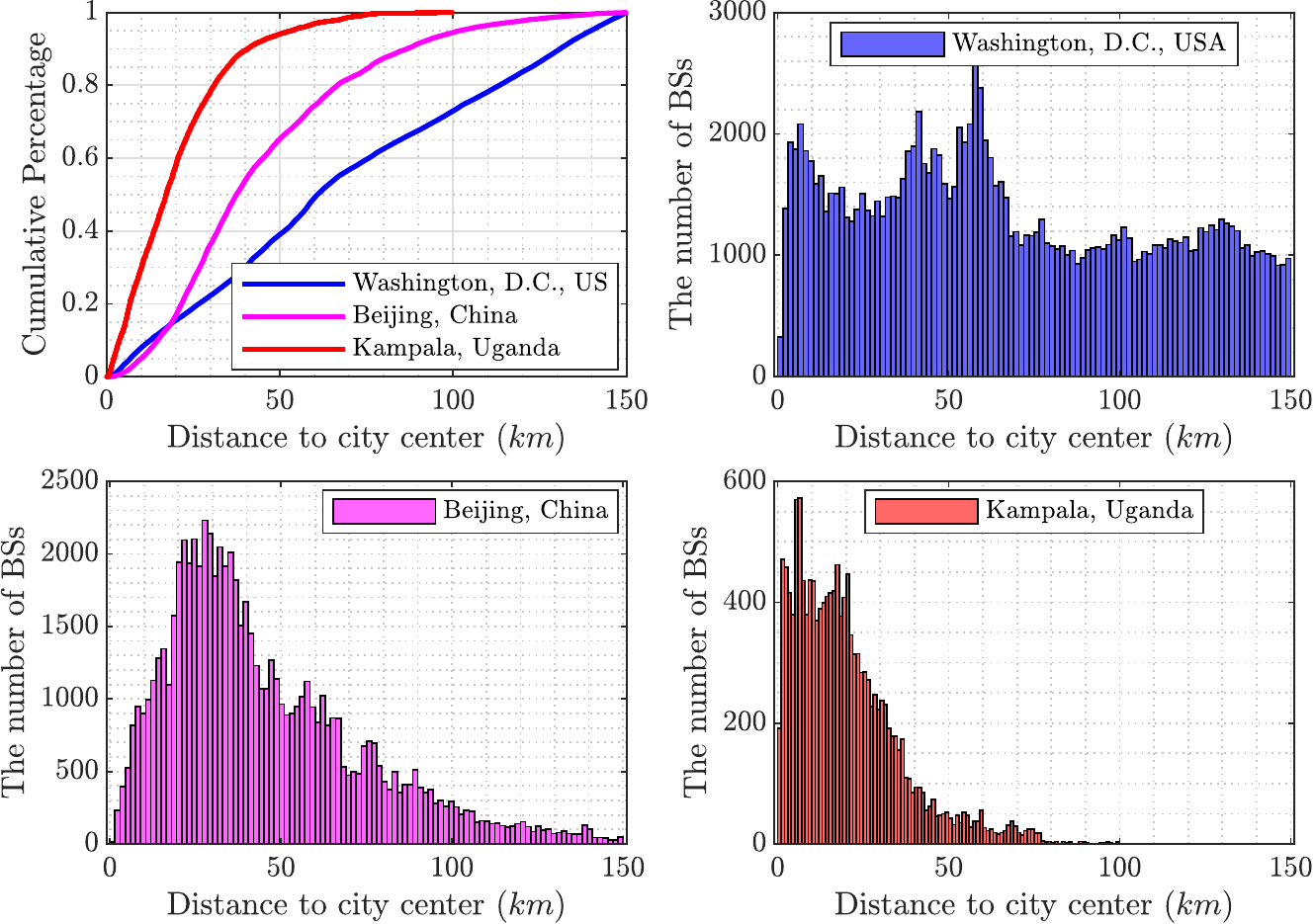}
\caption{Distribution of distances from each BS to the city center: Three case studies for representative cities.}
\label{cdf_pdf}
\end{figure}

By plotting the distribution of base stations (BSs) worldwide in Fig. \ref{bs_distribution}, one can easily and intuitively observe that many areas in the map have no or limited number of BSs, especially in Africa, Central Asia,  Northern and South America. Also, developed countries have better connectivity than developing and least developed countries, indicating the digital divide exists on a country-wise level. Besides, the digital divide also exists on an area-wise level inside a country. This phenomenon can be quantitatively reflected in Fig. \ref{cdf_pdf}, which shows the distribution of distances from each BS to the nearest city center. Three different cities, i.e., Washington, D.C., Beijing, and Kampala, are selected and treated as representative cities. The upper-left sub-figure in Fig. \ref{cdf_pdf} illustrates the cumulative probability distribution of distances from BSs to the city center for the selected three cities. The other three sub-figures represent the histograms of distances. From Fig. \ref{cdf_pdf}, it is direct to derive the following key points:
\begin{itemize}
\item The more developed the city, the less BSs are located in the city center.
\item The number of BSs near the city center is higher than that far away from the city center, regardless of the level of urbanization. For example, about $95\%$ of the BSs are located less than 50 kilometers from Kampala's city center; however, this number is only $40\%$ for Washington, D.C. These results validate the area-wise digital divide.
\end{itemize}

The obstructions causing the digital divide for the four billion of unconnected and under-connected people are sophisticated and region-specific, which can be classified into the technological and non-technological categories. 

The technological obstruction mainly refers to the lack of an efficient solution to construct telecommunication infrastructure to connect those unconnected. As the manufacturing cost of mobile handsets has continuously been falling over the last decade, mobile handsets are much affordable in recent years, which might not be a bottleneck problem for the connection at all \cite{5960025}. On the other hand, the construction and deployment plans relying on optical fibers and base stations for urban and populous areas might not be economically feasible for remote and rural areas, albeit with technological feasibility. Although in most countries, building infrastructure to cover the underdeveloped areas is a legal duty for telecommunication operators to run business, the lack of economic incentive dismisses their motivation, resulting in poor coverage over these underdeveloped areas. Furthermore, it is generally even more difficult and costly to build and maintain the telecommunication infrastructure by traditional techniques in underdeveloped areas, due to the lacks of legacy telecommunication infrastructure and electricity facility, adverse terrain, altitude, and other natural conditions. In the 6G era, new communication technologies and novel network architectures must be applied in conjunction with new business models to overcome the technological obstructions that prevent the four billion from global connectivity \cite{8877348}.

Apart from the technological obstruction, the non-technological obstructions are various, including the lacks of trust and content relevance, language barrier, Internet censorship, and low computer literacy. Apart from Internet censorship that is a supply-side issue, most non-technological obstructions are the issues with the demand side \cite{5960025}. In other words, the residents in the remote areas have no need or motivation to be connected. Although exchanging information is one of the basic needs of human beings, exchanging information can be realized in various ways without involving telecommunications.


\section*{Big Communications: The Solution}
To overcome the aforementioned obstructions and achieve the goal of global and ubiquitous connectivity covering the remaining four billion unconnected/under-connected people, we propose and detail a novel solution termed BigCom in this article. BigCom is proposed for achieving the following missions in the 2030s:
\begin{itemize}
\item Connect the unconnected population over the globe, especially those underrepresented groups with economic, geometric, and demographic disadvantages.
\item Democratize the benefits of ICT over the globe and close the digital divide step by step.
\item Promote the global and ubiquitous connectivity as the cornerstone of the United Nations' SDGs.
\end{itemize}

\begin{figure}[!t]
\centering
\includegraphics[width=3.5in]{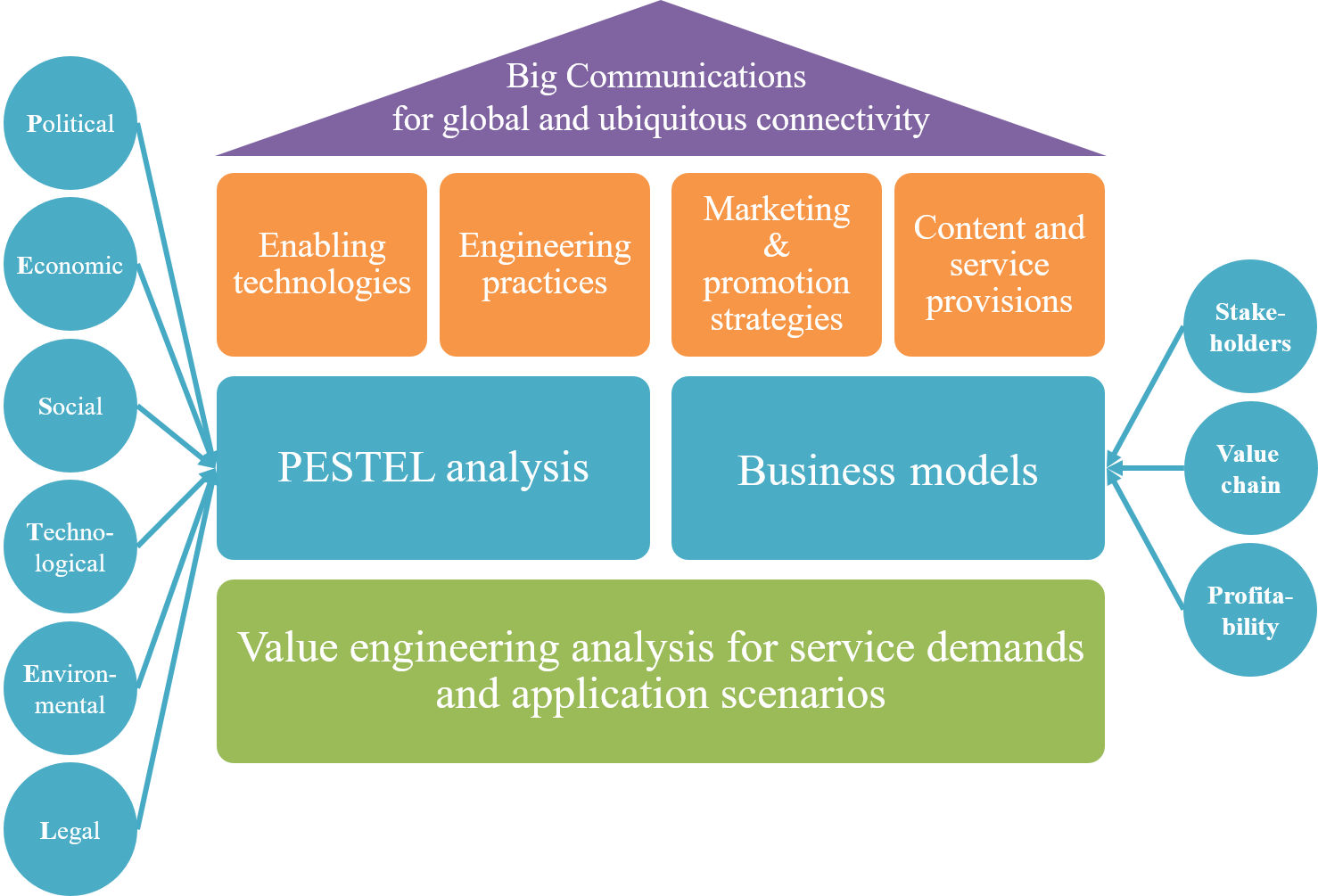}
\caption{Complete framework of BigCom for achieving global and ubiquitous connectivity and democratizing the benefits of ICT.}
\label{framework}
\end{figure}

Meanwhile, it should be noted that BigCom is not only a solution dedicated to solving the technical problems but also a comprehensive framework analyzing the service demands, socio-economic feasibility, and technological enablers, as well as their long-term implications for specific areas. Based on value engineering analysis, PESTEL analysis (PESTEL: an acronym that stands for \textbf{P}olitical, \textbf{E}conomic, \textbf{S}ocial, \textbf{T}echnological, \textbf{E}nvironmental, and \textbf{L}egal factors) and various business models, BigCom can be formed as a bottom-up framework as shown in Fig. \ref{framework} to help connect people living in underdeveloped regions with limited infrastructure and financial resources.

To provide effective and region-specific solutions under the framework of BigCom, researchers and engineers must first resort to value engineering analysis to assess the service demands and application scenarios for unconnected/under-connected people. The value engineering analysis constitutes the foundation of BigCom and is carried out to understand served groups from three perspectives:
\begin{itemize}
\item What are the demanded communication and data services when connected?
\item What are the special needs and peculiarities for the service provisions?
\item What is the minimum quality of service (QoS) that should be guaranteed? 
\end{itemize}

Generally, the outcomes of the value engineering analysis are region-specific, and the corresponding actions should be taken case by case. However, there exists certain universality among the collected surveys for different far-flung, rural, and underdeveloped areas, which is summarized and briefly discussed in the following paragraphs.

First, the habitual residence of people who are unconnected or under-connected decides that most of the communication demands are confined within a local and clustered zone \cite{9139048}. The demanded contents for communication and data services are highly associated with local affairs and daily life. Occasionally, there are some data demands through national and international connections, which are generated mainly for business purposes, instead of recreational purposes. 

The traits of communication and data demands of under-developed areas imply that the direct communication distance between two devices would be much closer than the case in urban and dense areas, and the node mobility in under-developed areas would be limited and even negligible. Also, considering the low-rate nature of service demands in these underdeveloped areas, the network transmission capacity does not need to be excessive. Still, the connection reliability is of paramount importance \cite{8082583}.

On the other hand, some case studies have also shown that once a region has been well connected and the residents get used to the telecommunication services, they might even demand more services, rendering a much higher network throughput. This indicates that when deploying the basic telecommunication infrastructure for underdeveloped areas, one should always keep a futuristic mind and remain a high degree of flexibility and scalability \cite{matinmikko2017micro}.

Apart from the aforementioned technological traits, considering the economic disadvantage of the users in underdeveloped areas, they are more sensitive to the communication cost than those living in developed areas. Affordability should be enhanced to promote the penetration of telecommunications in underdeveloped areas and connect more people so as to yield the scale effect for cost reduction \cite{9139048}. The affordability advantage in underdeveloped areas can be gained by trading customization off. 

In addition, both flexibility and affordability can also be reinforced by the flexible spectrum management in far-flung areas, as the spectrum occupation is generally sparse in these areas. The richness of available spectrum allows much more flexible and inexpensive use of spectral resources, because a number of constraints for mitigating co-channel interference can be lifted and various complex techniques for enhancing spectral efficiency are not required.

\begin{figure}[!t]
\centering
\includegraphics[width=3.5in]{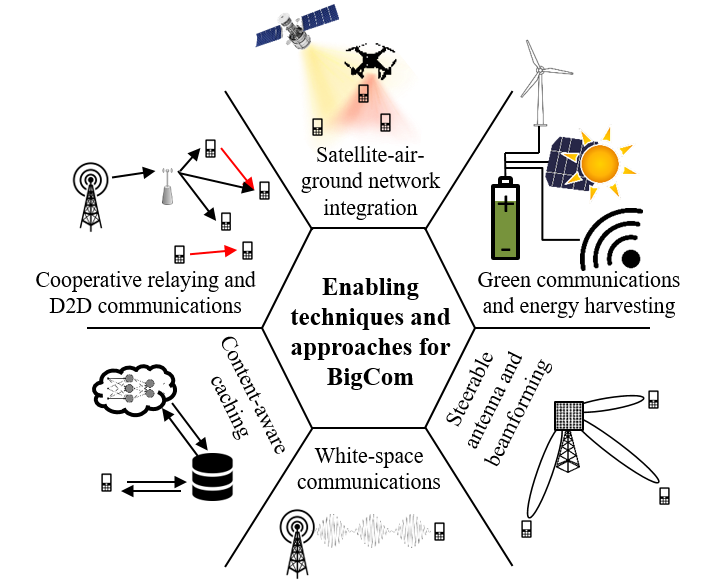}
\caption{Techniques and approaches enabling BigCom.}
\label{tech}
\end{figure}

Meanwhile, as a systematic project, the value engineering analyses frequently uncover the dependence of reliable connectivity on reliable power supply in underdeveloped areas. Lacking power infrastructure and intermittent power supply is one of the main causes for discontinuous communications in underdeveloped areas \cite{8082583}, and higher priority should be given to the energy efficiency of telecommunication networks by a limited energy budget.

\section*{Enabling Technologies of Big Communications}
Depending on different service demands and application scenarios, there exist a number of techniques and approaches in the toolbox of BigCom that can be jointly applied to connect the remaining four billion unconnected/under-connected people living in underdeveloped regions. We pictorially illustrate these techniques and approaches in Fig. \ref{tech} and introduce each of them with the corresponding application scenarios in the following paragraphs.

Based on the existing telecommunication infrastructure, e.g., core networks, optical cables, machine rooms, and base stations, cooperative relaying and device-to-device (D2D) communications would be two efficient physical-layer techniques for coverage expansion \cite{8344837}. Both techniques can work in a complementary manner to expand network coverage and enhance communication services without constructing new infrastructure. They are in particular efficient for satisfying clustered communication and data service demands in proximity.

Apart from the terrestrial telecommunication infrastructure, the space and airborne telecommunication infrastructure should also be integrated to provide reliable coverage over far-flung areas \cite{9177315}. In the context of 6G, conventional satellite communications are expected to be greatly enhanced by drones equipped with radio transceivers and processing modules (a.k.a. aerial base stations). As a direct result of the satellite-air-ground network integration, basic communication and data services can also be provided to the dead spots that cannot be covered solely by terrestrial radios. Benefiting from the high mobility and adjustable positions of drones, such integrated networks can be deployed in a dynamical manner. Therefore, as added bonuses, networking flexibility and scalability can also be considerably improved by this integration.

In addition, because rich spectral resources are available in underdeveloped areas, high-quality frequency bands and large bandwidths can be employed for large and reliable coverage. Among a number of candidate frequency bands, because of the unlicensed status and a good penetration ability, television white space (TVWS) spectrum (470-790 MHz in Europe) has been regarded as a promising solution for achieving wide and frugal broadband connectivity \cite{9139048}. The coverage radius of TVWS spectrum is typically 30 km and can be easily extended to 100 km if required. Meanwhile, multiple available frequency bands also allow performing subcarrier permutation among different cooperative hops to harvest frequency diversity gains and lead to better end-to-end communication performance.

To improve energy efficiency for underdeveloped areas with a limited energy budget for communications, steerable antenna arrays with beamforming functionality play a crucial role \cite{8877348}. This steerable architecture takes advantage of the low mobility of communication nodes in underdeveloped areas and can accurately track the geometric locations of nodes by a small amount of signaling overhead. To reduce the dependence on unreliable power grids, green communications and diverse energy harvesting techniques can be used to provide renewable energy for telecommunication infrastructure and equipment in underdeveloped regions. 

In higher layers, edge computing and artificial intelligence are applicable for facilitating content-aware caching for clustered communication and data services in rural areas. Because of the similar and highly predictable demanded contents, when the spectral and energy resources are sufficient, edge servers can pre-load contents that would likely be requested by users. Content-aware caching provides an opportunistic optimization mechanism for allocating communication and computing resources over time.

\section*{Suggestions for Implementing Big Communications}
According to the summary of service demands, application scenarios, and the enabling technologies for BigCom, we come up with several suggestions for implementing BigCom in practice as follows.

\subsection{Content providers}
Content providers are the key to solve the underlying issue of the digital divide. They should provide justifiable reasons to connect the unconnected people living in underdeveloped areas, which create motivation in technological and administrative innovations. Although somebody might argue that this is a `chicken or the egg' dilemma, it is clear that even with full-fledged telecommunication infrastructure, the implementation of BigCom will not succeed if the unconnected people living in underdeveloped areas need to pay extra but find no reason to be connected. 

Thus, rather than passively waiting for new niche markets and business opportunities, content providers are supposed to actively create new demands for those living in underdeveloped areas. Some helpful experience could be learned from the rapid digitalization of China in the past decades \cite{zhang2019china}. The content providers in China proposed a series of innovative digital services, including e-payment, online ordering and delivery, and peer-to-peer product supply, which created a large number of profitable niche markets. Because these innovative digital services greatly facilitate daily life and benefit the locals, the penetration rates of smart phones and the relevant wireless data services in China have surged to a record high in the past decade.

\subsection{Mobile network operators}
\begin{figure}[!t]
\centering
\includegraphics[width=3.2in]{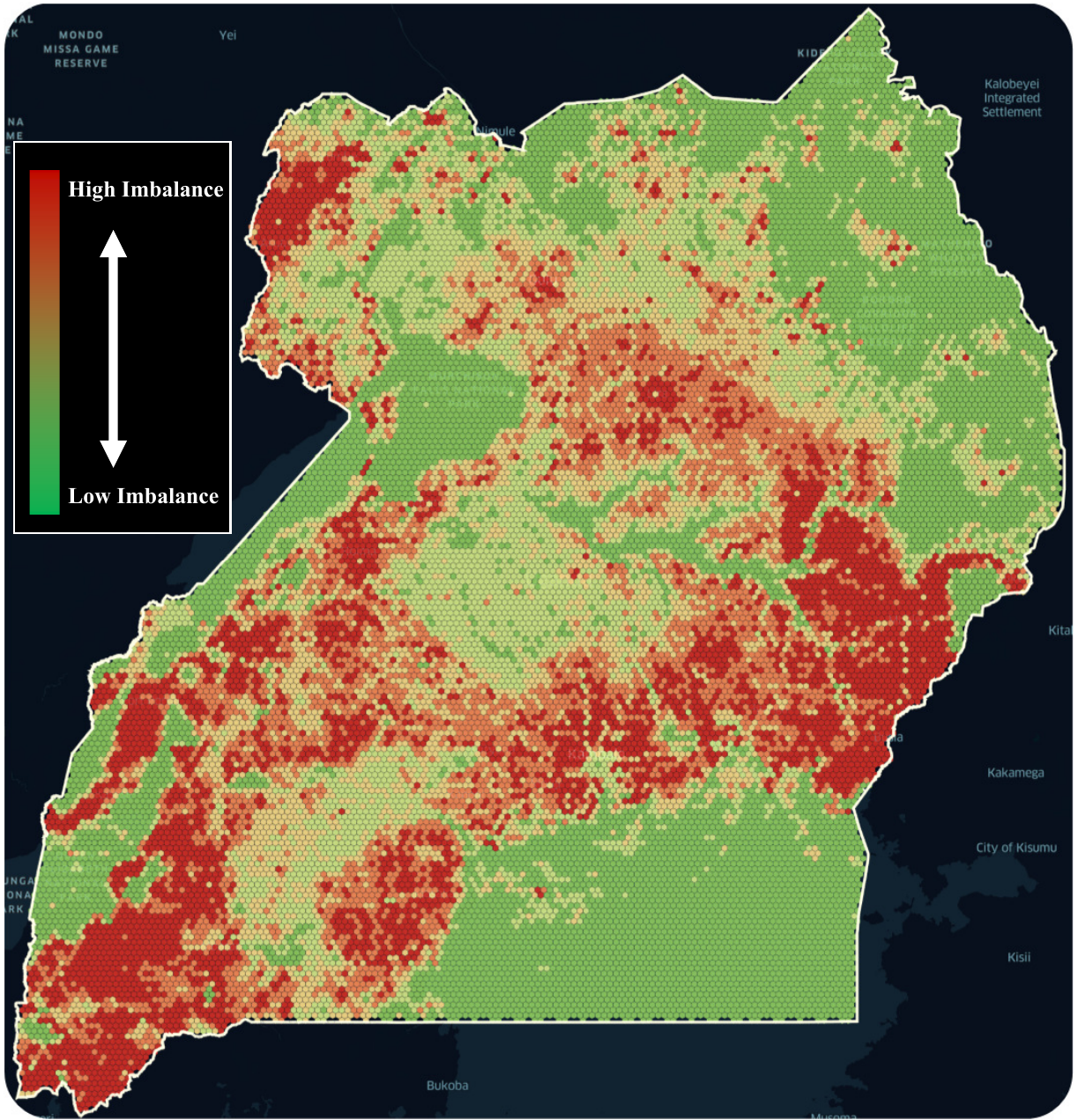}
\caption{Connectivity imbalance visualization for Uganda, a developing country in the middle of Africa. The whole country is spitted into many small areas, and each area is approximately $4~\mathrm{km}^2$, and  the imbalance index calculates a ratio of population to the number of BSs for each area.}
\label{uganda}
\end{figure}

The mobile network operators should fully release artificial intelligence's technological advantages and detect cell coverage holes based on cross-domain data fusion. For example, as shown in Fig. \ref{uganda}, by considering both the population data and the BS data, mobile network operators can easily find out which places are less covered and quantify the degree of lack of telecommunication resources. Aside from identifying coverage holes, operators should also resort to some cost-effective ways to cover rural areas, such as integrating heterogeneous telecommunication resources, including BSs, satellites, drones, and earth-orbiting balloons.

Apart from the resource integration under conventional business models, novel business models should also be found by mobile network operators to connect the unconnected people living in far-flung and rural areas. Currently, the idea of `who benefits who pays' has come into attention, which results in the concept of micro-operators \cite{matinmikko2017micro}. A micro-operator is both the service provider and the end user. However, the self-sufficient telecommunication services supported by micro-operators will compromise the vested interest of mobile network operators, and therefore how these novel business models can coalesce into conventional business models still remains an open problem.

\subsection{Governments}
Governments, in particular those of undeveloped countries, should recognize that accessing the Internet is the key to making innovation, improving education, and reducing poverty. Thus, governments might provide certain incentives to content providers and mobile network operators in the initial stage of telecommunication constructions. The potential incentives include but are not limited to flexible regulations of administration, tax reduction, direct financial supports, land resources, and spectral resources. These incentives will not only attract investment for telecommunication infrastructure but also reduce the costs of construction, maintenance, and operation of telecommunication networks with a small number of users.

Balancing fairness and efficiency is an eternal topic in policy making; governments should, however, be aware of the fact that it is unrealistic in the short term to provide comparable data services in rural areas as in urban areas. Such temporary imbalance should be tolerated, leading to different construction objectives and telecommunication policies in rural and urban areas. Therefore, governments might not expect to close the digital divide by one attempt but reduce the divide step by step and finally close it. In this regard, flexible policies and rich spectral resources should be leveraged as the advantages in far-flung and rural areas to facilitate basic and accessible data services for the unconnected people.

\section*{Concluding Remarks}
In this article, we analyzed the obstructions preventing four billion people from proper connections and proposed a novel framework termed BigCom for the purpose of realizing global and ubiquitous connectivity in the 6G era. BigCom is a comprehensive and hierarchical framework aiming at identifying, mitigating, and solving both technological and non-technological issues impeding linking the last remaining four billion people in far-flung, rural, and underdeveloped areas. With the support of BigCom for 6G, it is highly expected that the global and ubiquitous connectivity will not become the privilege of residents in dense and urban areas but also cover the four billion people living in underdeveloped areas. As a result, the benefits of ICT would be democratized among the globe so as to help with the realizations of the United Nations' SDGs and promote the progress of the entire human society.

\bibliographystyle{IEEEtran}
\bibliography{bib}

\begin{IEEEbiographynophoto}{Shuping Dang} [M'18] (shuping.dang@kaust.edu.sa) received D.Phil in Engineering Science from University of Oxford and is now a Postdoctoral Fellow with the Computer, Electrical and Mathematical Science and Engineering (CEMSE) Division, King Abdullah University of Science and Technology (KAUST).
\end{IEEEbiographynophoto}
\begin{IEEEbiographynophoto}{Chuanting Zhang}[M'18] (chuanting.zhang@kaust.edu.sa) received PhD in Communication and Information Systems from Shandong University and is now a Postdoctoral Fellow with the CEMSE Division, KAUST.
\end{IEEEbiographynophoto}
\begin{IEEEbiographynophoto}{Basem Shihada}[SM'12] (basem.shihada@kaust.edu.sa) received his PhD in Computer Science from University of Waterloo and is now an associate \& founding professor in the CEMSE Division at KAUST.
\end{IEEEbiographynophoto}
\begin{IEEEbiographynophoto}{Mohamed-Slim Alouini}[F'09] (slim.alouini@kaust.edu.sa) received the Ph.D. degree in Electrical Engineering from the California Institute of Technology (Caltech) and served as a faculty member at the University of Minnesota and then at the Texas A\&M University, before joining KAUST as a Professor of Electrical Engineering in 2009. 
\end{IEEEbiographynophoto}

\end{document}